\documentclass[acmsmall,screen]{acmart}
\usepackage{tikz}
\usepackage{mathtools}
\usetikzlibrary{arrows.meta}
\usepackage{lipsum}
\usepackage[linesnumbered,ruled]{algorithm2e}
\usepackage[font=itshape]{quoting}
\AtBeginDocument{%
  \providecommand\BibTeX{{%
    \normalfont B\kern-0.5em{\scshape i\kern-0.25em b}\kern-0.8em\TeX}}}

\acmJournal{JACM}
\acmVolume{0}
\acmNumber{0}
\acmArticle{0}
\acmMonth{8}

\newtheorem{myDef}{Definition}
\newtheorem{myPro}{Proof}
\begin{document}

\title{The Robustness Verification of Linear Sound Quantum Classifiers}

\author{SuBonan}
\affiliation{
  \institution{Shandong University}
  \streetaddress{72 Binhai Road, Jimo}
  \city{Qingdao}
  \country{China}}
\email{bokkenasu@163.com}

\renewcommand{\shortauthors}{SuBonan}

\begin{abstract}
I present a quick and sound method for the robustness 
verification of a sort of quantum classifiers who 
are \emph{Linear Sound}.\;Since quantum machine learning has been put into practice in relevant fields and 
\emph{Linear Sound Property, LSP} is a pervasive property, the method could be universally applied. I implemented 
my method with a \emph{Quantum Convolutional Neural Network, QCNN} using \emph{MindQuantum, Huawei} and successfully
verified its robustness when classifying \emph{MNIST} dataset.
\end{abstract}

\ccsdesc[100]{Computing methodologies~Quantum machine learning; Robustness verification}
\ccsdesc[100]{Theory of computation~Quantum computation}

\keywords{Quantum Classifier, Quantum Neural Network, Robustness verification, Linear Sound Property}

\maketitle

\section{Background}
In this section, I would like to recall some basic conceptes of quantum mechanics and quantum computation models to help
amateur readers have a comprehensive understanding of my idea.\\
\subsection{Basic concepts and notations}
One reasonable way for us to view the objects is to consider in two separated aspects: \emph{the abstract information (data)} and \emph{its physical carrier}.
Taking classical computer as an example, we use the mathematical notation 0 and 1 (bits) to abstract the data while its physical carrier is \emph{electrical level}. \\
According to the hypothesis of quantum mechanics, \emph{quantum bits (qubits)} are playing the same role to represent data.
A single qubit could be expressed by a normalized complex vector $|\psi\rangle=\begin{pmatrix}a \\ b\end{pmatrix}$ satisfying $||a||^2+||b||^2=1$ where $|\psi\rangle$ is the \emph{Dirac notation}.
The most significant difference between bits and qubits is that bits are discrete while qubits could be any \emph{superposition} between $|0\rangle=\begin{pmatrix}1 \\ 0\end{pmatrix}$ and $|0\rangle=\begin{pmatrix}0 \\ 1\end{pmatrix}$.
There are various elaborate physical devices who are able to carry qubits like \emph{Ion trap} \cite{kielpinski2002architecture} or \emph{Photon}\cite{wilk2007single}, but we don't need to pay much attention to the physcal carrier when it comes to theoretical analysis.\\
The two base quantum states $\{|0\rangle, |1\rangle\}$ is an orthonormal base of a 2-dimentional \emph{Hilbert space} while in general cases, a quantum system containing n qubits could be represented by a $2^n$-dimensional normalized complex vector.
Such states are often referred as \emph{Pure state}.\\
\subsection{Quantum gate and density operator}
Under the ideal condition, a \emph{Quantum gate} is mathematically represented by unitary matrices $U$ on $\mathcal{H}$, i.e. $U^\dagger U=UU^\dagger = I$.
Just like bits, the evolution of qubits could be described by a sequence of quantum gates, which is a \emph{Quantum circuit}.
The result we apply a quantum gate is:
\begin{equation}
  |\psi_{out}\rangle=U|\psi_{in}\rangle
\end{equation}
However, in real world, the state of a quantum system may not be completely detected and can be
considered as a \emph{Mixed state}: $\{p_i,|\psi_i\rangle\}$ where $\forall i. p_i\in[0, 1],|\psi_i\rangle\in\mathcal{H}$ and $\sum_{i}p_i=1$.
Its intuitive meaning is that the quantum system is at $|\psi_i\rangle$ with probability $p_i$. 
Another mathematical way to represent quantum state is \emph{Density operator}:
\begin{equation}
\rho=\sum_ip_i|\psi_i\rangle\langle\psi_i|
\end{equation}
where $\langle\psi_i|=|\psi_i\rangle^\dagger$.
The density operator is a Hermitian positive semidefinite matrix and has several properties: 
\begin{itemize}
    \item [$\bullet$]{$trace(\rho)=1$}
    \item [$\bullet$]{$trace(\rho^2)\leq 1$}
    \item [$\bullet$]{$trace(\rho^2)=1$ if and only of $\rho$ is a pure state ($\rho=|\psi\rangle\langle\psi|$)}
\end{itemize}
Thanks to the density operator, we could explain quantum circuit (or quantum evolution) as a \emph{Super-operator} $\mathcal{E}$:
\begin{equation}
  \rho_{out}=\mathcal{E}_{circuit}(\rho_i)=U_nU_{n-1}...U_2U_1\rho U_1^\dagger U_2^\dagger...U_{n-1}^\dagger U_n^\dagger
\end{equation}

where $\{U_1,U_2,...,U_{n-1},U_n\}$ is the sequence of quantum gate (quantum circuit).
\subsection{Measurement}
The only way we observe or obtain information from a quantum system is to \emph{Measure} it. 
The mathematical model for the measurement is a set of matrics on $\mathcal{H}$: $\{M_1,M_2,...,M_m\}$ satisfying $\sum_{i=1}^m M_i^\dagger M_i=I$.
The measurement could be viewed as a mapping from quantum states to classical bits: the result of the measurement is k with probability $p_k=tr(M_k\rho M_k^\dagger)$.
Ideally, if we repeat the measurement, we could get a probability distribution over $\{1, 2, ... ,m\}$.\\
However, the measurement will change the quantum states depending on its result.
After measurement, the system will collaspe into the new state:$\rho=\frac{M_k\rho M_k^\dagger}{tr(M_k\rho M_k^\dagger)}$ if the result of measurement is k, which makes it extremely difficult to get the distribution precisely.
One reasonable and effective method is to obtrain enough copies of initial quantum states.
\subsection{summary}
To put it briefly, the quantum states could be classifed into \emph{Pure states} or \emph{Mixed states} and are represented by \emph{Density operator} or \emph{Dirac notations}.
The reason why quantum computing functs much faster than classical computing is that a single qubit is able to carry infinte information since it's continuous rather than discrete.
Nevertheless, the observation of quantum system will destroy it, which requires us to design quantum algorithm ingeniously. 

\section{Formalizations}
In this section, I'm going to formalize the definition of quantum classifiers and its relevant concepts.
Besides, the robustness verification problem ought to be formally illustrated for the convenience of later analysis.
\begin{myDef}
  \emph{A Quantum Classifier} is a mapping $D(\mathcal{H})\rightarrow \mathcal{C}$
  where $\mathcal{H}$ is a given Hilbert space and $D(\mathcal{H})$ refers
  to the set of all (mixed) quantum states on $\mathcal{H}$. $\mathcal{C}$ stands for
  the set of classes we are interested in.
\end{myDef}
Just like conventional classifiers and neural networks, it's intuitive and formal that a
\emph{Quantum Neural Network, QNN} (which can function as a quantum classifier) 
consts of two primary parts: 
\begin{itemize}
      \item [$\bullet$]{\emph{A Parameterized Quantum Circuit}}
      \item [$\bullet$]{\emph{A Classification Policy}}
  \end{itemize}
\subsection{Parameterized quantum circuit}
\begin{myDef}
  \emph{A Parametric Quantum Circuit} is a quantum circuit composed of \emph{Paramtetirc Unitary Gates}
  (like $R_x(\theta),R_y(\theta),R_z(\theta)$) and other basic quantum gates(like $H,X,CNOT$).
\end{myDef}
Futhermore, the circuit could be divided into \emph{Encoder} (optional) and \emph{Ansatz}. 
The former circuit is designed to encoder classical data into quantum states, therefore
its parameters depend on the data set and can not be trained while the latter circuit's
parameters are all trainable.\\
The \emph{Encoder} circuit is optional in that the input data could be classical in practice
as well as quantum states and sometimes the circuit is substituted by elaborate physical devices
which are more efficient.\\
The \emph{Ansatz} circuit is mathematically modelled as a quantum \emph{Super-operator}
$\mathcal{E}_\theta:\rho\rightarrow\rho'$ where $\theta$ is a set of free parameters
that can be tuned. The paradigm of ansatz is not absolute and lots of efforts have been
put into the design and analysis of ansatz.

\subsection{Classification policy}
\begin{myDef}
  \emph{A Classification Policy} $\mathcal{P}$ is a partial function $\{p_1,p_2,...,p_m\}\rightarrow\mathcal{C}$,
  where $\{p_1,p_2,...,p_m\}$ is a probability distribution over $\{1,2,...,m\}$ .
\end{myDef}
I chose the probability distribution as the input of classification policy instead of the original measurement result as a matter of convenience. 
In practive, one can apply statistics methods (like \emph{Hypothesis testing}) to approximate the distribution.\\
One of the most primitive but useful classification policy is to perform
measuremrnt in the standard computational base, which is, to choose 
$|0\rangle\langle 0|,|1\rangle\langle 1|,|2\rangle\langle 2|,...,|c\rangle\langle c|$
as the collection of measurement operators where c is the number of classes: $c = |\mathcal{C}|$.
Then $\mathcal{P}(p_1,p_2,...,p_c)=\mathop{\arg\max}\limits_{p_i,1\leq i\leq c}i$, which means that we select the most probable measurement result as the classification output.
\\
Intuitively, the classification policy mentioned above is to classify quantum state by its probablity amplitude. 
Moreover, there are many other classification policies. 
For example, \emph{Hybrid Quantum-Classical Neural Networks} \cite{liu2021hybrid} are gaining popularity these days in that the model allows us to choose different statistics in the result of quantum measurement
as the input data for a classical neural network, which indicates a more complicated classification policy.
\subsection{An illustrative example: Quantum Convolutional Neural Network, QCNN}
Let me give an instance of the definitions mentioned above: \emph{QCNN}\cite{cong2019quantum}, which is successfully applied in image recognition.\\
\begin{figure}[h]
  \centering
  \includegraphics[width=0.8\textwidth]{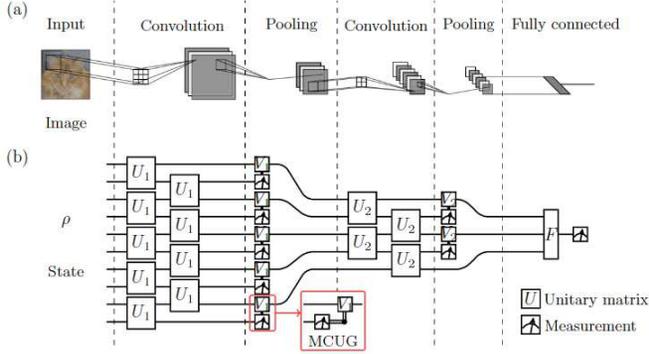}
  \caption{Comparison between CNN and QCNN}
\end{figure}

Figure 1 indicates the relationship between classical convolutional neural network and quantum convolutional neural network.
Similar to CNN, the QCNN contains \emph{Convolutional layer} and \emph{Pooling layer}, which two constitute the \emph{Ansatz} circuit.
The \emph{Encoder} circuit is neglacted in the figure while in practice, I chose the most straightforward way: \emph{Amplitude encoding}. \\
The convolutional layer is made up of two-qubits parametric gates applying on every adjacent qubits, which results in global entanglement.
The pooling layer consists of \emph{Measurement control unitary gate, MCUG}, which gate will funct if and only if the outcome of measurement under computational base is 1.\\
The classification policy could be extremely simple: 
\begin{equation}
\mathcal{P}(p_0,p_1)=\begin{cases}0&p_0>p_1+\varepsilon \\ 1&p_1>p_0+\varepsilon \\ Unknown&|p_0-p_1|\leq\varepsilon\end{cases}
\end{equation}
Intuitively we are just employing binary classification according to the measurement on the last single qubit.
One may realize that there is an area where the class is "Unknown", in fact, we can improve the accuracy of the model by deserting some opaque cases who lies near the boundary between two classes.\\
The choices of ansatz parametric gate and more analysis about QCNN have been discussed before \cite{hur2022quantum}, more details about the implementation will also be referred to in later sections.

\subsection{Robustness verification problem}
Just as Gehr et al. proposed\cite{gehr2018ai2}, we also need to consider how robust a quantum classifier is against quantum noise. \\
Intuitively speaking, we hope our classifer is robust enough to classify two "very close" quantum states into the same class so that the slight disturbance will not influence the classification.
In image recognition, the "closeness" should reflect the perceptual similarity between two images, so it is essential to give a formal and reasonable definition of the "closeness" in quantum mechanics.\\
Fortunately, we already have the \emph{Fidelity} to quantify the "closeness" and the uncertainty caused by quantum noise\cite{myerson2008high}:
\begin{equation}
Fidelity(\rho,\sigma)=tr(\sqrt{\sqrt{\rho}\sigma\sqrt{\rho}})^2
\end{equation}

where $\rho=\sum_{i}p_i|\psi_i\rangle\langle\psi_i|$ and $\sigma=\sum_{k}q_k|\phi_k\rangle\langle\phi_k|$ are two density operators representing two different quantum systems
and $\sqrt{\rho}=\sum_{i}\sqrt{p_i}|\psi_i\rangle\langle\psi_i|$.
There are also other ways to estimate the uncertainty like \emph{Super fidelity}\cite{puchala2011experimentally}, we would proof that it works just similar to fidelity. 
So here in this paper, I chose the fidelity:

\begin{myDef}
  \emph{The Distance} between two quantum states $\rho, \sigma$ is defined as:
  \begin{equation}
    Dis(\rho, \sigma)=1-Fidelity(\rho,\sigma)=1-tr(\sqrt{\sqrt{\rho}\sigma\sqrt{\rho}})^2
  \end{equation}
\end{myDef}

After the definition of distance, we are ready to give the definition for the robustness of classifiers: 

\begin{myDef}
  A quantum classifier is \emph{$\delta$-Robust} if and only if for every quantum states $\rho$ in training dataset,
 we have $\forall \sigma\in\mathcal{H}, Dis(\rho,\sigma)\leq\delta\Rightarrow\rho,\sigma$ are classified the same.
\end{myDef}

I defined the robustness by training dataset here, while in some pratical situations, we can discuss the robustness with respect to validation dataset or any correctly classified dataset.\\
Now the robust verification problem is quite natural in that we hope to find the $\varepsilon$ and ensure that the classifier is $\varepsilon$-robust as well.

\section{Verification}
This section is the core part of this paper.
Before introducing proofs and equations, I prefer to state the centeral idea here, which is intuitive but inspiring:
\begin{quoting}
  The essence of classification is to divide the state space into different parts. \\
  What's more, for linear sound classifiers, they're just dividing the whole space into convex sets with hyperplanes.
\end{quoting}

In later sections I would illustrate the idea in a formalized mathematical way.
\subsection{General bloch vectors}
\emph{The Bloch vector} is a geometric tool to represent quantum state.
For a single qubit, the state vector could always be written as $|\psi\rangle=cos\frac{\theta}{2}|0\rangle+e^{i\varphi}sin\frac{\theta}{2}$ which forms a one-to-one mapping to the vectors$(sin\theta cos\varphi, sin\theta sin\varphi, cos\theta)$ on the 3-D unit \emph{Bloch Sphere}.\\
When it comes to multi-qubits, the case is a bit more complicated.
Since \emph{The General Pauli Matrices} forms an orthonormal base of Hilbert space, we can also find a correspondence between density operator and vectors.
For every valid density operator $\rho$, the equation below always holds:

\begin{equation}
\rho=\frac{1}{2^n}(I^{\otimes n}+\overrightarrow{v}\cdot\overrightarrow{\sigma})
\end{equation}

where n is the number of qubits, $\overrightarrow{v}\in\mathbb{R}^{4^n-1}$ is the corresponding vector and $\overrightarrow{\sigma}=\{I, X, Y, Z\}^{\otimes n}-I^{\otimes}$ stands for the set of n-qubits general pauli matrices.
For $n=1$, $\overrightarrow{\sigma}=(X, Y, Z)$ and for $n=2$, $\overrightarrow{\sigma}=(I\otimes X, I\otimes Y,I\otimes Z, X\otimes I,X\otimes X,X\otimes Y,X\otimes Z, Y\otimes I,Y\otimes X,Y\otimes Y,Y\otimes Z,Z\otimes I,Z\otimes X,Z\otimes Y,Z\otimes Z)$.\\
The general pauli matrices forms an orthonormal base due to the self-evident property:
\begin{equation}
  \forall i,j. tr(\sigma_i\cdot\sigma_j)=2^n\delta_{ij} \; and\;
  \forall i. tr(\sigma_i)=0
\end{equation}

where $\delta_{ij}$ is \emph{Kronecker Symbol}: $\delta_{ij}=\begin{cases}1&i=j \\ 0 & i\neq j\end{cases}$.
In this paper, I amended the pauli base by adding an coefficient: $\overrightarrow{\sigma_{norm}}=\sqrt{2^n-1}\overrightarrow{\sigma}$,
thanks to which the following statement can be proved:
\begin{equation}
||\overrightarrow{v}||\leq 1\;and\;||\overrightarrow{v}||=1\Leftrightarrow \rho\;is\;a\;pure\;state.
\end{equation}
where $\rho=\frac{1}{2^n}(I^{\otimes n}+\overrightarrow{v}\cdot\overrightarrow{\sigma_{norm}})$.
\begin{myPro}
  According to the principle of quantum mechanics, we know that for all density operator $\rho$, $tr(\rho^2)\leq 1$, and the equality holds if and only if $\rho$ is a pure state. \\
  Besides, we have:
  \begin{equation}
    \begin{split}
      tr(\rho^2) &=tr(\frac{1}{4^n}(I^{\otimes n}+2\cdot\overrightarrow{v}\cdot\overrightarrow{\sigma_{norm}}+(\overrightarrow{v}\cdot\overrightarrow{\sigma_{norm}})\cdot(\overrightarrow{v}\cdot\overrightarrow{\sigma_{norm}}))) \\
      &=\frac{1}{4^n}(2^n+0+tr((\overrightarrow{v}\cdot\overrightarrow{\sigma_{norm}})\cdot(\overrightarrow{v}\cdot\overrightarrow{\sigma_{norm}})))\\
      &=\frac{1}{4^n}(2^n+(4^n-2^n)\overrightarrow{v}\cdot\overrightarrow{v})\\
      &=\frac{1+(2^n-1)||\overrightarrow{v}||^2}{2^n}  
    \end{split}
  \end{equation}
  where the last line is because $tr(\sigma_{normi}\sigma_{normj})=(4^n-2^n)\delta_{ij}$\\
  now it's obvious that $tr(\rho^2)<1 \Leftrightarrow ||\overrightarrow{v}||^2<1$ and $tr(\rho^2)=1\Leftrightarrow ||\overrightarrow{v}||^2=1$.
\end{myPro}
From now on, I will use $\overrightarrow{\sigma}$ for substitute of $\overrightarrow{\sigma_{norm}}$.
\subsection{Neighbourhood}
Considering the robustness verification problem, what we need to ensure is that the whole neighbourhood of one central quantum states are classified the same.
let $\omega(\rho)=\{\sigma | Dis(\rho, \sigma) \leq\delta\}$ denotes the $\delta$-neighbourhood of $\rho$.\\
We can interprete it in a geometric way. 
For two pure states $\rho=|\psi\rangle\langle\psi|,\sigma=|\phi\rangle\langle\phi|$, noticing that:
\begin{equation}
  \begin{split}
    \because Fidelity(\rho,\sigma) &=tr(|\psi\rangle\langle\psi|\phi\rangle\langle\phi|)\\
    &=\frac{1}{2^n}(1+(2^n-1)\overrightarrow{v_{\rho}}\cdot\overrightarrow{v_{\phi}})\\
    \therefore Dis(\rho,\sigma)\leq\delta &\Leftrightarrow Fidelty(\rho,\sigma)\geq 1-\delta \\
    &\Leftrightarrow \overrightarrow{v_{\rho}}\cdot\overrightarrow{v_{\phi}}\geq \frac{2^n(1-\delta)-1}{2^n-1} 
  \end{split}
\end{equation}
Now we can reconsider the $\delta$-neighbourhood, since $||\overrightarrow{v_{\rho}}||=||\overrightarrow{v_{\phi}}||=1$(see Proof 1.), let $\omega^\#(\rho)=\{\sigma|cos\theta\geq\frac{2^n(1-\delta) - 1}{2^n-1}\}$ 
where $\theta$ is the angle between $\overrightarrow{v_{\rho}}$ and $\overrightarrow{v_{\phi}}$.
We can prove that $\omega(\rho)=\omega^\#(\rho)$:

\begin{myPro}
  noticing that $\rho$ is a pure state: $\rho=|\psi\rangle\langle\psi|$
  \begin{equation}
    \begin{split}
      \because Fidelity(\rho,\sigma) &=tr(\sqrt{|\psi\rangle\langle\psi|\sigma|\psi\rangle\langle\psi|})^2 \\
      &=\langle\psi|\sigma|\psi\rangle(tr(\sqrt{|\psi\rangle\langle\psi|}))^2\\
      &=\langle\psi|\sigma|\psi\rangle\\
      &=tr(\rho\sigma) \\
      \therefore Dis(\rho,\sigma)\leq\delta &\Leftrightarrow Fidelity(\rho,\sigma)\geq 1-\delta \\
      &\Leftrightarrow tr(\rho\sigma)\geq 1-\delta\\
      &\Leftrightarrow \frac{1+(2^n-1)\overrightarrow{v_{\rho}}\cdot\overrightarrow{v_{\sigma}}}{2^n}\geq 1-\delta \\
      &\Leftrightarrow \overrightarrow{v_{\rho}}\cdot\overrightarrow{v_{\sigma}}\geq\frac{2^n(1-\delta)-1}{2^n-1} \\
      \therefore \forall \sigma,\sigma\in\omega(\rho) &\Leftrightarrow \sigma\in\omega^\#(\rho)\\
      \therefore \omega(\rho) &=\omega^\#(\rho)
    \end{split}
  \end{equation}
\end{myPro}

Intuitively, $\omega^\#(\rho)$ is somehow like a "cylinder", the figure below shows a blue geometric object containing $\varepsilon$-neighbourhood around $\rho$ when there is only one qubit.
From now on, I use $\omega(\rho)$ to denote both $\omega(\rho)$ and $\omega^\#(\rho)$.
\begin{center}
\begin{tikzpicture}[line cap=round, line join=round, >=Triangle]
  \clip(-2.19,-2.49) rectangle (2.66,2.58);
  \draw [shift={(0,0)}, lightgray, fill, fill opacity=0.1] (0,0) -- (96.4: 0.4) arc (96.4:123.6:0.4) -- cycle;
  \draw(0,0) circle (2cm);
  \draw [rotate around={0.:(0.,0.)},dash pattern=on 3pt off 3pt] (0,0) ellipse (2cm and 0.9cm);
 
  \draw [->, color=red] (0,0)-- (-0.80,1.2);
  \draw [color=blue] (0,0)-- (-1.13,0.67);
  \draw [color=blue] (0,0)-- (-0.14,1.25);
  \draw [rotate around = {30 : (-0.7, 1.07)}, color=blue](-0.7,1.07) ellipse (0.6cm and 0.45cm);
  \draw [->] (0,0) -- (0,2);
  \draw [->] (0,0) -- (-0.81,-0.79);
  \draw [->] (0,0) -- (2,0);
  \draw (-0.43,0.9) node[anchor=north west] {$\theta$};
  \draw (-1.01,-0.72) node[anchor=north west] {$\mathbf {\hat{x}}$};
  \draw (2.07,0.3) node[anchor=north west] {$\mathbf {\hat{y}}$};
  \draw (-0.5,2.6) node[anchor=north west] {$\mathbf {\hat{z}=|0\rangle}$};
  \draw (-0.4,-2) node[anchor=north west] {$-\mathbf {\hat{z}=|1\rangle}$};
  \draw (-0.8,1.6) node[anchor=north west] {$\rho$};
  \scriptsize
  \draw [fill] (0,0) circle (1.5pt);
\end{tikzpicture}
\end{center}

\subsection{The Linear Sound Property, LSP}
Before introducing the core concept LSP, let's take a look on the initial idea again:
\begin{quoting}
  The essence of classification is to divide the state space into different parts. \\
  What's more, for linear sound classifiers, they're just dividing the whole space into convex sets with hyperplanes.
\end{quoting}

For example, consider the classification policy mentioned in section 3.2 when there is only one qubit:
\begin{equation}
\mathcal{P}(p_0,p_1)=\begin{cases}0&p_0>p_1+\varepsilon \\ 1&p_1>p_0+\varepsilon \\ Unknown&|p_0-p_1|\leq\varepsilon\end{cases}
\end{equation}

As the figure shown below, we can understand it in a geometric way:
if the bloch vector's projection on z-axis is above the green circular column, then the state will be classified zero while those below the circular column will lead to class 1.
\begin{center}
  \begin{tikzpicture}[line cap=round, line join=round, >=Triangle]
    \clip(-2.5,-2.49) rectangle (2.66,2.58);
    \draw(0,0) circle (2cm);
    \draw [->] (0,0) -- (0,2);
    \draw [->] (0,0) -- (-0.81,-0.79);
    \draw [->] (0,0) -- (2,0);
    \draw [rotate around={0.:(0.,0.)},dash pattern=on 3pt off 3pt] (0,0) ellipse (2cm and 0.9cm);  
    \draw (-1.01,-0.72) node[anchor=north west] {$\mathbf {\hat{x}}$};
    \draw (2.07,0.3) node[anchor=north west] {$\mathbf {\hat{y}}$};
    \draw (-0.5,2.6) node[anchor=north west] {$\mathbf {\hat{z}=|0\rangle}$};
    \draw (-0.4,-2) node[anchor=north west] {$-\mathbf {\hat{z}=|1\rangle}$};
    \filldraw [green, fill opacity=0.1, rotate around={0.:(0.,0.)}] (0,0.2) ellipse (2cm and 0.9cm);  
    \filldraw [green, fill opacity=0.1, rotate around={0.:(0.,0.)}] (0,-0.2) ellipse (2cm and 0.9cm); 
    \draw [green](-2, 0.2) -- (-2, -0.2);
    \draw [green](2, 0.2) -- (2, -0.2);
    \draw [](-2, 0.2) -- (-2.2, 0.2);
    \draw [](-2, -0.2) -- (-2.2, -0.2);
    \draw [->] (-2.1, 0) -- (-2.1, 0.2);
    \draw [->] (-2.1, 0) -- (-2.1, -0.2);
    \draw (-2.09, -0.2) node[anchor=south east] {$2\varepsilon$};

    \draw (0.3, 1.2) node[anchor=south west] {$|\psi\rangle$ into class 0};
    \draw [->] (0,0) -- (0.3, 1.2);
    \draw (-0.8, -1.7) node[anchor=north] {$|\phi\rangle$ into class 1};
    \draw [->] (0,0) -- (-0.8, -1.7);
    \scriptsize
    \draw [fill] (0,0) circle (1.5pt);
  \end{tikzpicture}
  \end{center}

It's time that I should give a formal definition of this kind of classifiers:
\begin{myDef}
  We say a classifier have the \emph{Linear Sound Property, LSP}, if and only if it satisfies:
  \begin{itemize}
    \setlength{\itemsep}{2pt}
    \item {if states $\rho_1,\rho_2$ are classified into the same class, then for any $p\in[0,1]$, $p\rho_1+(1-p)\rho_2$ is classified into the same class too.}
    \item {if state $\rho=\frac{1}{2^n}(I^{\otimes n}+\overrightarrow{v}\cdot\overrightarrow{\sigma})$ is classified into $c$, then for any $l\geq||\overrightarrow{v}||$, 
    $\rho'=\frac{1}{2^n}(I^{\otimes n}+\frac{\overrightarrow{v}}{l}\cdot\overrightarrow{\sigma})$ is classified into $c$ (if $\rho'$ can be classifed).}
  \end{itemize}
\end{myDef}

We can prove that an obvious example of Linear Sound Classifier is the combination of any quantum circuit and the classification policy mentioned above.
\begin{myPro}
  For the circuit part, it's quite self-evident that after applying a unitary gate $U$, those who are on the line between bloch vector $\overrightarrow{v_1},\overrightarrow{v_2}$ 
  are still on the line between vector $\overrightarrow{v_1'},\overrightarrow{v_2'}$ and that
  those who are on the extension line of bloch vector $\overrightarrow{v}$ are still on the same direction:
  \begin{equation}
    \begin{split}
      \forall \rho_1,\rho_2 &,\; p\in[0, 1] \\
      \rho_1 &=\frac{1}{2^n}(I^{\otimes n}+\overrightarrow{v_1}\cdot\overrightarrow{\sigma}),\;
      \rho_2 =\frac{1}{2^n}(I^{\otimes n}+\overrightarrow{v_2}\cdot\overrightarrow{\sigma}) \\
      \rho_3 &= p\rho_1+(1-p)\rho_2=\frac{1}{2^n}(I^{\otimes n}+(p\overrightarrow{v_1}+(1-p)\overrightarrow{v_2})\cdot\overrightarrow{\sigma})\\
    U\rho_1 U^\dagger &= \frac{1}{2^n}(I^{\otimes n} + U\overrightarrow{v_1}\cdot\overrightarrow{\sigma} U^\dagger),\; U\rho_2 U^\dagger = \frac{1}{2^n}(I^{\otimes n} + U\overrightarrow{v_2}\cdot\overrightarrow{\sigma} U^\dagger) \\
    U\rho_3 U^\dagger &= \frac{1}{2^n}(I^{\otimes n}+U(p\overrightarrow{v_1}+(1-p)\overrightarrow{v_2})\cdot\overrightarrow{\sigma})U^\dagger\\
    &=\frac{1}{2^n}(I^{\otimes} + p U\overrightarrow{v_1}\cdot\overrightarrow{\sigma}U^\dagger + (1-p)U\overrightarrow{v_2}\cdot\overrightarrow{\sigma}U^\dagger) \\
    let\;\overrightarrow{v_1}'\cdot\overrightarrow{\sigma} &=U\overrightarrow{v_1}\cdot\overrightarrow{\sigma} U^\dagger,\;
    \overrightarrow{v_2}'\cdot\overrightarrow{\sigma} =U\overrightarrow{v_2}\cdot\overrightarrow{\sigma} U^\dagger \\
    \overrightarrow{v_3}'\cdot\overrightarrow{\sigma} &=pU\overrightarrow{v_1}\cdot\overrightarrow{\sigma}U^\dagger + (1-p)U\overrightarrow{v_2}\cdot\overrightarrow{\sigma}U^\dagger \\
    &=[p\overrightarrow{v_1'}+(1-p)\overrightarrow{v_2'}]\cdot\overrightarrow{\sigma} \\
    \because \overrightarrow{\sigma}\; &is\;an\;orthonormal\;base\\
    \therefore \overrightarrow{v_3'} &=p\overrightarrow{v_1'}+(1-p)\overrightarrow{v_2'}\\
    What's\;more:\\
    \forall \rho' &=\frac{1}{2^n}(I^{\otimes n}+\frac{\overrightarrow{v_1}}{l}\cdot\overrightarrow{\sigma}) \\
    U\rho'U^\dagger &=\frac{1}{2^n}(I^{\otimes n}+\frac{U\overrightarrow{v_1}\cdot\overrightarrow{\sigma}U^\dagger}{l}) \\
    let\;\overrightarrow{v''}\cdot\overrightarrow{\sigma} &=\frac{U\overrightarrow{v_1}\cdot\overrightarrow{\sigma}U^\dagger}{l}=\frac{\overrightarrow{v_1'}\cdot\overrightarrow{\sigma}}{l}=\frac{\overrightarrow{v_1'}}{l}\cdot\overrightarrow{\sigma}\\
    then\;\overrightarrow{v''} &=\frac{\overrightarrow{v_1'}}{l}\\
  \end{split}
  \end{equation}
  So we have proved that $\overrightarrow{v_3}=p\overrightarrow{v_1}+(1-p)\overrightarrow{v_2}\Rightarrow \overrightarrow{v_3'}=p\overrightarrow{v_1'}+(1-p)\overrightarrow{v_2'}$ and $\overrightarrow{v'}=\frac{\overrightarrow{v_1}}{l}\Rightarrow\overrightarrow{v''}=\frac{\overrightarrow{v_1'}}{l}$. \\
  Intuitively, unitary gates function as rotation operators on the bloch vectors so that they won't change the relative position. \\
  {\noindent} \rule[0pt]{\textwidth}{0.05em}
  For the classification policy part, we need to consider the measurement of the first qubit where the measurement operators are $|0\rangle\langle 0|\otimes I^{\otimes (n-1)},|1\rangle\langle 1|\otimes I^{\otimes (n-1)}$.
  let $tr_{/0}(\rho)$ stand for the partial trace $tr_{1, 2, ...,n-1}(\rho)$:
  \begin{equation}
    \begin{split}
      \because p_0 &=tr(|0\rangle\langle 0|_{0}\rho|0\rangle\langle 0|_0),\;p_1=tr(|1\rangle\langle 1|_{0}\rho|1\rangle\langle 1|_0) \\
      \therefore p_0 &=\langle 0|tr_{/0}(\rho)|0\rangle,\;p_1=\langle 1|tr_{/0}(\rho)|1\rangle\\
    \end{split}
  \end{equation}
  If we take a look at the elements in $\overrightarrow{\sigma}=\sqrt{2^n-1}(\{I,X,Y,Z\}^{\otimes n}-I^{\otimes n})$, we would realize that $tr_{/0}(\sigma_i)\neq 0$ if and only if $\sigma_i=\sqrt{2^n-1}X\otimes I^{\otimes (n-1)}$ or $\sqrt{2^n-1}Y\otimes I^{\otimes (n-1)}$ or $\sqrt{2^n-1}Z\otimes I^{\otimes (n-1)}$ 
  in that $tr(X)=tr(Y)=tr(Z)=0$. So we have:
  \begin{equation}
    \begin{split}
      p_0 &=\frac{1}{2^n}(2^{n-1}+2^{n-1}\sqrt{2^n-1}[v_0\langle 0|X|0\rangle + v_1\langle 0|Y|0\rangle+v_2\langle 0|Z|0\rangle]) \\
        &= \frac{1+\sqrt{2^n-1}v_2}{2} \\
      p_1 &=\frac{1-\sqrt{2^n-1}v_2}{2}  \\
      So\; p_0 > p_1 &\Leftrightarrow v_2>0 \\
      p_0-p_1\geq\varepsilon &\Leftrightarrow v_2>\frac{\varepsilon}{\sqrt{2^n-1}} \\
      p_1-p_0\geq\varepsilon &\Leftrightarrow v_2<-\frac{\varepsilon}{\sqrt{2^n-1}} \\
    \end{split}
  \end{equation}
  The equivalence relation reveals the initial idea again, the essence of classification is just to judge whether one component of the bloch vector is greater than a $\varepsilon$-related value.
  So the proof is obvious: $v_2>\frac{\varepsilon}{\sqrt{2^n-1}}\; and\;u_2>\frac{\varepsilon}{\sqrt{2^n-1}}\Rightarrow\forall p\in[0,1],pv_2+(1-p)u_2>\frac{\varepsilon}{\sqrt{2^n-1}}$.
  And $\frac{v_2}{l}$ maintains the sign of $v_2$.\\
  Especailly when n=1, the classification policy is equivalent to $z>\varepsilon$, shown by the figure at the begining of this section 4.3.
\end{myPro}

The proof of \emph{LSP} is always the core part of the verification.
In fact I have proved that both quantum gate and the measurement maintain LSP, so what really matters is the classification policy $\mathcal{P}$.
An typical classification policy that is not linear sound is to put the result of the measurement into a classical neural network, forming an quantum-classical hybrid neural network.
What's more, LSP actually guarantees that the class sets are convex in essence.
\subsection{Verification}
Now we are ready to verify the neighbourhood.
Intuitively, what we need to find is the minimum angle between the central state vector $\overrightarrow{v_\rho}$ and the hyper plane $v_2=\frac{\varepsilon}{\sqrt{2^n-1}}$ if $\rho$ is classified as 0.
Technically and mathematically, we need to calculate:
\begin{equation}
  \begin{split}
    cos\theta_{min}=\mathop{\max}\limits_{||\overrightarrow{u}||=1,u_2=\frac{\varepsilon}{\sqrt{2^n-1}}}\overrightarrow{v_\rho}\cdot\overrightarrow{u}
  \end{split}
\end{equation}
We can prove that for any vector $\overrightarrow{u}$ satisfying $\overrightarrow{u}\cdot\overrightarrow{v_\rho}>cos\theta_{min}$, then we have $u_2>\frac{\varepsilon}{\sqrt{2^n-1}}$ meaning that the state is classified into 0.
The opposite side $u_2<-\frac{\varepsilon}{\sqrt{2^n-1}}$ is actually the same.
\begin{myPro}
  Let $\overrightarrow{v_\rho}=(v_0,v_1,..,v_{2^n-1})$, $v_2>\frac{\varepsilon}{\sqrt{2^n-1}}$. Consider the set of vectors: $L(\varepsilon)=\{\overrightarrow{x}|x_2=\frac{\varepsilon}{\sqrt{2^n-1}},||\overrightarrow{x}||=1\}$, which describes "a layer of vectors". 
  We need to prove that $cos\theta_{min}=\mathop{\max}\limits_{||\overrightarrow{x}||=1,\overrightarrow{x}\in L(\varepsilon)}=\overrightarrow{x}\cdot\overrightarrow{v_\rho}$ is monotone increasing with $t=\frac{\varepsilon}{\sqrt{2^n-1}}$,
  So that if given a unit vector $\overrightarrow{u}$ satisfying $\overrightarrow{u}\cdot\overrightarrow{v_\rho}>cos\theta_{min}$ then we have $\overrightarrow{u}\in L(>\varepsilon)$, in other words, $u_2>\frac{\varepsilon}{\sqrt{2^n-1}}$.
  \begin{equation}
    \begin{split}
      & cos\theta=v_0x_0+v_1x_1+v_2t+...+v_{2^n-1}x_{2^n-1} \\
      & \because \begin{cases}v_0^2+v_1^2+v_3^2+...+v_{2^n-1}^2=1-v_2^2\\ x_0^2+x_1^2+x_3^2+...+x_{2^n-1}^2=1-t^2\end{cases} \\
      & \qquad(v_0x_0+v_1x_1+v_3x_3+...+v_{2^n-1}x_{2^n-1})^2\leq (v_0^2+v_1^2+v_3^2+...+v_{2^n-1}^2)(x_0^2+x_1^2+x_3^2+...+x_{2^n-1}^2) \\
      & \qquad and\;the\;equality\;could\;be\;achieved\;when\;\frac{v_0}{x_0}=\frac{v_1}{x_1}=...=\frac{v_{2^n-1}}{x_{2^n-1}} \\
      & \therefore cos\theta_{min}=v_2t+\sqrt{(1-v_2^2)(1-t^2)} \\
      & \qquad \frac{dcos\theta_{min}}{dt}=v_2-\sqrt{\frac{1-v_2^2}{1-t^2}}t \\
      & \because v_2>t \\
      & \therefore \frac{dcos\theta_{min}}{dt}>v_2-t>0
    \end{split}
  \end{equation}
  So $cos\theta_{min}$ is monotone increasing with $\varepsilon$.
\end{myPro}
I will visualize the proof in 3D case:
\begin{center}
  \begin{tikzpicture}[line cap=round, line join=round, >=Triangle]
    \clip(-2.5,-2.49) rectangle (2.66,2.58);
    \draw(0,0) circle (2cm);
    \draw [->] (0,0) -- (0,2);
    \draw [->] (0,0) -- (-0.81,-0.79);
    \draw [->] (0,0) -- (2,0);
    \draw [rotate around={0.:(0.,0.)},dash pattern=on 3pt off 3pt] (0,0) ellipse (2cm and 0.9cm);  
    \draw (-1.01,-0.72) node[anchor=north west] {$\mathbf {\hat{x}}$};
    \draw (2.07,0.3) node[anchor=north west] {$\mathbf {\hat{y}}$};
    \draw (-0.5,2.6) node[anchor=north west] {$\mathbf {\hat{z}=|0\rangle}$};
    \draw (-0.4,-2) node[anchor=north west] {$-\mathbf {\hat{z}=|1\rangle}$};
    \filldraw [green, fill opacity=0.1, rotate around={0.:(0.,0.)}] (0,1) ellipse (1.67cm and 0.6cm);  
    
    \draw [->, color=red] (0,0) -- (0.7,1.87);
    \draw (0.7, 1.87) node[anchor=south west] {$\overrightarrow{v_\rho}$};
    \draw [->, color=blue] (0,0) -- (1.632, 1.1);
    \draw (1.632, 1.1) node[anchor = south west] {$\overrightarrow{v_{min}}$};
    \draw [shift={(0,0)}, lightgray, fill, fill opacity=0.1] (0,0) -- (33.98: 0.4) arc (33.98:69.47:0.4) -- cycle;
    \draw (0.1, 0.1) node[anchor = south west] {$\theta_{min}$};
    \draw [->] (0,0) -- (1.25, 1.57);
    \draw (1.25, 1.57) node [anchor = south west] {$\overrightarrow{u}$};
    \draw [shift={(0,0)}, lightgray, fill, fill opacity=0.1] (0,0) -- (51.42: 0.7) arc (51.42:69.47:0.7) -- cycle;
    \draw (0.2, 0.5) node[anchor = south west] {$\theta$};
    \draw [->, color=green] (1.8, 0.55) -- (1.8, 1.1);
    \draw [->, color=green] (1.8, 0.55) -- (1.8, 0);
    \draw [] (1.632, 1.1) -- (1.9, 1.1);
    \draw (1.8, 0.55) node [anchor = west] {$\varepsilon$};
    \draw [->, color=blue,dash pattern=on 3pt off 3pt] (0, 1.1) -- (1.632, 1.1);
    \draw [->, color=red,dash pattern=on 2pt off 2pt] (0, 1.1) -- (0.7, 1.1);
    \draw [dash pattern = on 2pt off 2pt] (0.7, 1.1) -- (0.7, 1.87);
    \scriptsize
    \draw [fill] (0,0) circle (1.5pt);
  \end{tikzpicture}
\end{center}
The figure above shows that, for a fixed central state $\overrightarrow{v_\rho}$ and a hyper plane $z=\varepsilon$, 
if $\overrightarrow{v_\rho}$ is above the plane, then for any given state $\overrightarrow{u}$, if the angle between $\overrightarrow{v_\rho}$ and $\overrightarrow{u}$ is smaller than any angle between $\overrightarrow{v_\rho}$ and the vectors on the bound of the plane($\theta<\theta_{min}$), then $\overrightarrow{u}$ would be above the plane too.
What's more, the project of $\overrightarrow{v_{min}}$ and $\overrightarrow{v_\rho}$ on the surface $z=\varepsilon$ will lie on the same line (the red and blue dotted horizontal vector) because of the condition of the equality. 
\begin{center}
  \begin{tikzpicture}[line cap=round, line join=round, >=Triangle]
    \clip(-2.5,-2.49) rectangle (2.66,2.58);
    \draw(0,0) circle (2cm);
    \draw [->] (0,0) -- (0,2);
    \draw [->] (0,0) -- (-0.81,-0.79);
    \draw [->] (0,0) -- (2,0);
    \draw [rotate around={0.:(0.,0.)},dash pattern=on 3pt off 3pt] (0,0) ellipse (2cm and 0.9cm);  
    \draw (-1.01,-0.72) node[anchor=north west] {$\mathbf {\hat{x}}$};
    \draw (2.07,0.3) node[anchor=north west] {$\mathbf {\hat{y}}$};
    \draw (-0.5,2.6) node[anchor=north west] {$\mathbf {\hat{z}=|0\rangle}$};
    \draw (-0.4,-2) node[anchor=north west] {$-\mathbf {\hat{z}=|1\rangle}$};
    \filldraw [green, fill opacity=0.1, rotate around={0.:(0.,0.)}] (0,0.2) ellipse (2cm and 0.9cm);  
    \filldraw [green, fill opacity=0.1, rotate around={0.:(0.,0.)}] (0,-0.2) ellipse (2cm and 0.9cm); 
    \draw [green](-2, 0.2) -- (-2, -0.2);
    \draw [green](2, 0.2) -- (2, -0.2);
    \draw [](-2, 0.2) -- (-2.2, 0.2);
    \draw [](-2, -0.2) -- (-2.2, -0.2);
    \draw [->] (-2.1, 0) -- (-2.1, 0.2);
    \draw [->] (-2.1, 0) -- (-2.1, -0.2);
    \draw (-2.09, -0.2) node[anchor=south east] {$2\varepsilon$};
    \draw [->, color=red] (0,0) -- (1.87,0.7);
    \draw (1.87, 0.7) node[anchor=south west] {$\overrightarrow{v_\rho}$};
    \draw [color=blue, rotate around={110.5:(1.683,0.63)}] (1.683,0.63) ellipse (0.5cm and 0.2cm); 
    
    \draw [->, color=blue] (0,0) -- (1.86, 0.15);
    \draw [->, color=blue] (0,0) -- (1.5, 1.1);
    \draw [shift={(0,0)}, lightgray, fill, fill opacity=0.1] (0,0) -- (4.61: 0.4) arc (4.61:20.52:0.4) -- cycle;
    
    \scriptsize
    \draw [fill] (0,0) circle (1.5pt);
    \draw (0.3, -0.1) node[anchor=south west] {$\theta_{min}$};
  \end{tikzpicture}
\end{center}
The figure above indicates that the process of verification is actually to find the $cos\theta_{min}$ so that the neighbourhood "cylinder" will all be above the plane, so that the whole neighbourhood will be classified the same as the central state $\rho$,
which means that the $\varepsilon-$robustness is verified. \\
In summary, the core operation is to calculate $v_2=\frac{2p_0-1}{\sqrt{2^n-1}}$, $cos\theta_{min}=v_2\frac{\varepsilon}{\sqrt{2^n-1}}+\sqrt{(1-v_2^2)(1-\frac{\varepsilon^2}{2^n-1})}$ (see Proof 4.),
and then we can give a robust bound $\delta = 1-Fidelty=1-\frac{1+(2^n-1)cos\theta_{min}}{2^n}$.
\section{Implementation}
In this section, I'm going to illustrate the whole process of the experiment.
\subsection{QCNN and MNIST dataset}
\emph{MNIST database} (abbreviation for \emph{Modified National Institute of Standards and Technology database}) is a large database of handwritten digits that is commonly used for training various image processing systems. \\
Due to the limitation on the capability of classical computers to simulate quantum circuit, I filtered the dataset to do the binary classification only (left labels are mere 0s and 1s).
Besides, since I have chosen the amplitude encoder and to simulate 8 qubits, the maximum dimention of input data is $2^8=256$, 
I have reshaped the images into $16\times 16$ pixels with a simple algorithm.
For a image's data vector $(v_1,v_2,...,v_{256})$, the result of quantum amplitude encoder is $|\psi\rangle=\sum_{i=0}^{255}v_{i+1}|i\rangle$.
The circuit of the amplitude encoder is not complex and I have contributed it to the library of \emph{MindQuantum}. \\
The choice of the ansatz circuit is various, Tak Hur et al. have discussed the performance of different ansatz circuits\cite{hur2022quantum}.
In this paper, I chosed the most simple convolutional circuit shown below:

\begin{figure}[h]
  \centering
  \includegraphics[width=0.5\textwidth]{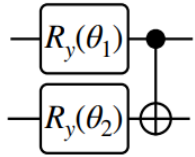}
  \caption{The convolutional circuit}
\end{figure}

while the pooling layer is just the combination of controlled $RZ$ and $RX$ gates.
The cost function is \emph{Softmax Cross Entropy Loss} and the network is trained by \emph{Adam optimizer}.
Such a simple QCNN is able to have a good enough performance: over $90\%$ accuracy of classfication on the test dataset.

\subsection{Process and results of the verification}
Although the proof before has comprehensively explained the verification process, I'm not tired of the repetition in pseudocode:
\begin{algorithm}[]
  \DontPrintSemicolon
    \SetAlgoLined
    \KwIn {A valid quantum state in training dataset $|\psi\rangle$, a well trained quantum circuit $c$, a accuracy tolerance $\varepsilon$}
    \KwOut {the result of classification, and the robust bound $\delta$}
    Apply the circuit $c$ on $|\psi\rangle$ to get $|\psi'\rangle$. \\
    Repeat the measurement of $|\psi'\rangle$ in computational basis for several times to get $p_0$. \\
    $v_2=\frac{2p_0-1}{\sqrt{2^n-1}}$. \\
    \eIf {$|p_0-p_1|\leq \varepsilon$}{
        panic "Can't be classified!".
    }{
      \eIf {$|p_0-p_1|>\varepsilon$}{
        $cos\theta_{min}=v_2\frac{\varepsilon}{\sqrt{2^n-1}}+\sqrt{(1-v_2^2)(1-\frac{\varepsilon^2}{2^n-1})}$ \\
        class = 0
      }{
        $cos\theta_{min}=-v_2\frac{\varepsilon}{\sqrt{2^n-1}}+\sqrt{(1-v_2^2)(1-\frac{\varepsilon^2}{2^n-1})}$ \\
        class = 1
      }
    }
    $\delta=1-\frac{1+(2^n-1)cos\theta_{min}}{2^n}$ \\
    return class, $\delta$.
    \caption{Verification of a specified quantum state in trainning dataset}
\end{algorithm}

The algorithm is extremely easy to understand and quite simple since the number of class is only two.
Extending the algorithm to the verification of classifiers who have more target classes is just to add some "If-else" cases and modify the way to calculate $cos\theta_{min}$ slightly. \\
It's self-evident that larger the discrepancy between $p_0$ and $p_1$ is, bigger $v_2$ is, larger robust bound we can get. \\
Table 1 and Table 2 below show a portion of the experiment's results, which suggests that for class 0, $v_2>0$ and $v_2<0$ for class 1.
Besides, for the same $p_0$ and $p_1$, when $\varepsilon$ is bigger, we have smaller $\theta_{min}$, bigger $cos\theta_{min}$ and smaller $\delta$.

\begin{table}[h]
  \begin{tabular}{ccccccc}
  \toprule
  \multicolumn{7}{c}{$\varepsilon=0$}                                                                                                     \\
  \hline
   & \multicolumn{1}{c}{$p_0$} & \multicolumn{1}{c}{$p_1$} & \multicolumn{1}{c}{$v_2$} & \multicolumn{1}{c}{$cos\theta_{min}$} & \multicolumn{1}{c}{$\delta$} &  \multicolumn{1}{c}{class} \\
  \midrule
   1     & $0.625719$ & $0.374281$ & $0.0157457$ & $0.999876$ &  $0.000123487$ & $0$ \\
   2     & $0.133918$ & $ 0.86608$ & $-0.0458499$ & $0.998948$ & $0.00104755$ & $1$ \\
   3     & $0.115384$ & $0.884616$ & $-0.0481711$ & $0.998839$ & $0.00115637$ & $1$ \\
   4     & $0.11719$  & $0.88281$  & $-0.0479449$ & $0.99885$  & $0.00114553$ & $1$ \\
   5     & $0.688041$ & $0.311959$ & $0.0235512$  & $0.999723$ & $0.000276284$ & $0$ \\
  \bottomrule
  \end{tabular}
  \caption{The result of verification when $\varepsilon = 0$}
\end{table}

\begin{table}[h]
  \begin{tabular}{ccccccc}
  \toprule
  \multicolumn{7}{c}{$\varepsilon=0.01$}                                                                                                     \\
  \hline
   & \multicolumn{1}{c}{$p_0$} & \multicolumn{1}{c}{$p_1$} & \multicolumn{1}{c}{$v_2$} & \multicolumn{1}{c}{$cos\theta_{min}$} & \multicolumn{1}{c}{$\delta$} &  \multicolumn{1}{c}{class} \\
  \midrule
   1     & $0.625719$ & $0.374281$ & $0.0157457$ & $0.999886$ &  $0.00011386$ & $0$ \\
   2     & $0.133918$ & $ 0.86608$ & $-0.0458499$ & $0.998977$ & $0.00101915$ & $1$ \\
   3     & $0.115384$ & $0.884616$ & $-0.0481711$ & $0.998869$ & $0.00112652$ & $1$ \\
   4     & $0.11719$  & $0.88281$  & $-0.0479449$ & $0.99888$  & $0.00111582$ & $1$ \\
   5     & $0.688041$ & $0.311959$ & $0.0235512$  & $0.999737$ & $0.000261789$ & $0$ \\
  \bottomrule
  \end{tabular}
  \caption{The result of verification when $\varepsilon = 0.01$}
\end{table}

\bibliographystyle{ACM-Reference-Format}
\bibliography{bibs}

\end{document}